\documentclass[11pt]{article}
\usepackage{geometry}             
\usepackage{graphicx}
\usepackage{amssymb}
\usepackage{amsmath}
\usepackage{epstopdf}
\usepackage{comment}
\usepackage{cite}
\usepackage{color}

\definecolor{brightpink}{rgb}{1.0, 0.0, 0.5}
\usepackage[hyperindex=true,
          pdfstartview=FitH,
          bookmarksnumbered=true,
          bookmarksopen=true,
          citecolor=blue,
          linkcolor=blue,
          colorlinks=true,
          unicode]{hyperref}

\newcommand{\be}{\begin{equation}}
\newcommand{\ee}{\end{equation}}
\newcommand{\bea}{\begin{eqnarray}}
\newcommand{\eea}{\end{eqnarray}}

\begin{document}
	
	\title{Hawking Evaporation of Einstein-Gauss-Bonnet AdS Black Holes in $D\geqslant 4$ dimensions}
	\author{
		Chen-Hao Wu${}^{1,}{}^{2,}{}^{4}$\thanks{{\em E-mail}:\href{mailto:chenhao\_wu@nuaa.edu.cn}
			{chenhao\_wu@nuaa.edu.cn}},  Ya-Peng Hu
		${}^{1,}{}^{4}$\thanks{{\em E-mail}:\href{mailto:huyp@nuaa.edu.cn}
		{huyp@nuaa.edu.cn}},  Hao Xu${}^{2,}{}^{3}$\thanks{{\em Corresponding author, E-mail}:\href{mailto:haoxu\_phys@163.com}{haoxu\_phys@163.com}}
		\vspace{5pt}\\
		\small ${}^{1}$College of Science, Nanjing University of Aeronautics and Astronautics,\\
		\small Nanjing, 210016, China\\
		\small  ${}^{2}$Center for Gravitation and Cosmology, College of Physical Science and Technology,\\
		\small Yangzhou University, Yangzhou, 225009, China\\
		\small  ${}^{3}$School of Aeronautics and Astronautics, Shanghai Jiao Tong University,\\
		\small Shanghai, 200240,  China\\
		\small ${}^{4}$MIIT Key Laboratory of Aerospace Information Materials and Physics,\\
		\small  Nanjing University of Aeronautics and Astronautics, Nanjing, 210016, China\\
	}

	\date{}
	\maketitle

\begin{abstract}

Einstein-Gauss-Bonnet theory is a string-generated gravity theory when approaching the low energy limit. By introducing the higher order curvature terms, this theory is supposed to help to solve the black hole singularity problem. In this work, we investigate the evaporation of the static spherically symmetric neutral AdS black holes in Einstein-Gauss-Bonnet gravity in various spacetime dimensions with both positive and negative couping constant $\alpha$. By summarizing the asymptotic behavior of the evaporation process, we find the lifetime of the black holes is dimensional dependent. For $\alpha>0$, in $D\geqslant6$ cases, the black holes will be completely evaporated in a finite time, which resembles the Schwarzschild-AdS case in Einstein gravity. While in $D=4,5$ cases, the black hole lifetime is always infinite, which means the black hole becomes a remnant in the late time. Remarkably, the cases of $\alpha>0, D=4,5$ will solve the terminal temperature divergent problem of the Schwarzschild-AdS case. For $\alpha<0$, in all dimensions, the black hole will always spend a finite time to a minimal mass corresponding to the smallest horizon radius $r_{min}=\sqrt{2|\alpha|}$ which coincide with an additional singularity. This implies that there may exist constraint conditions to the choice of coupling constant.

\end{abstract}

\section{Introduction}
In 1974, by applying quantum field theory in curved spacetime, Hawking showed that quantum mechanical effects allow black holes to emit particles, now known as Hawking radiation \cite{Hawking:1974rv,Hawking:1974sw}. Hawking radiation connects quantum physics and gravity, and it is viewed as a window to help us to understand the quantum gravity. Due to the radiation, the black hole will keep losing mass and thermal entropy, which leaves us the controversial information loss paradox \cite{Almheiri:2012rt,Hawking:2016sgy}. 

The spectrum of the Hawking radiation allows us to calculate the particles emission power and to estimate the lifetime of the black hole. For a static spherically symmetric black hole in four-dimensional asymptotically flat spacetime in Einstein gravity, its lifetime obeys the relationship $t\sim M_0^3$ associated with the initial black hole mass $M_0$ \cite{Page:1976df}, which means the black hole lifetime will be divergent if we take an infinite initial mass. A similar analysis can also be carried out in rotating black holes and charged black holes. A rotating black hole will lose its angular momentum much faster than its mass and eventually it becomes a Schwarzschild black hole and keeps going on its evaporation\cite{Page:1976ki}(unless the case with a lot of scalar particle species\cite{9710013,9801044}). A charged black hole will keep losing its charge and mass, however, the charge-over-mass ratio may increase and gets close to the extreme black hole, so the lifetime of these black holes may be extended by a huge factor, but finally, it also evolves to a Schwarzschild state\cite{Hiscock:1990ex,Xu:2019wak}.

The study on the evaporation of the black hole in asymptotically flat spacetime has been a fruitful area, while the evaporation in anti-de Sitter (AdS) spacetimes was, however, largely overlooked. This is due to the specific asymptotical behavior of the AdS spacetime. Although AdS has an infinite volume, the gravitational potential acts effectively like a finite confining box for the black hole, and the emitted massive particles reflect inward in a finite time thus preventing the energy loss. Thanks to the AdS/CFT correspondence \cite{Maldacena:1997re,Witten:1998qj,Witten:1998zw,Gubser:1998bc}, a great mass of attention has been paid on black holes in AdS spacetimes, especially AdS black hole thermodynamics\cite{Kubiznak:2014zwa}, and the evaporation problem of AdS black hole also comes back to our view in recent years. By imposing an absorbing AdS boundary condition \cite{Avis:1977yn,0804.0055,1304.6483,1307.1796} and considering only massless emitted particle, Page calculated the lifetime of static spherically symmetric AdS black hole in Einstein gravity \cite{Page:2015rxa}, and surprisingly, the lifetime did not become divergent with any arbitrarily large initial mass and the total time to evaporate from infinite mass to zero was finite, of the order $\ell^3$, where $\ell$ is the AdS radius. This is because although the black hole may have infinite initial mass, the temperature, and also the particles' emission power, are also divergent, thus the infinite amount of mass can be evaporated away in a finite time. The study on black holes evaporation in asymptotically AdS spacetimes have been extended to many different backgrounds, such as Lovelock gravity\cite{Xu:2019krv}, Conformal (Weyl) Gravity\cite{Xu:2018liy}, Ho\v{r}ava-Lifshitz gravity \cite{Xu:2020xsl}, and dRGT massive gravity\cite{Hou:2020yni}, etc.

From a modern perspective, Einstein gravity could just be an effective field theory in the low energy limit of unknown more fundamental theory. Although we have not known its correct and precise form, there are some attempts by physicists. Inspired by the quadratic term in heterotic string theory and six-dimensional Calabi-Yau compactifications of M-theory, the Gauss-Bonnet quadratic term was added into the Einstein-Hilbert action to analyze the properties of the effective gravity theory\cite{Boulware:1985wk,Lovelock:1971yv,Garraffo:2008hu,Callan:1988hs,Clifton:2011jh}. This gravity can be derived by string theories when approaching the low energy limit. Studies on the Gauss-Bonnet gravity and its generalizations have attracted lots of attention \cite{Cai:2001dz,Cvetic:2001bk,Torii:2005xu,Konoplya:2010vz,Xu:2013zea,Xu:2014tja,Xu:2015hba,Sun:2016til,Araneda:2016iiy,Konoplya:2017ymp,Konoplya:2017zwo}, however, this theory just exists in higher spacetime dimensions because the Gauss-Bonnet term is a topological invariant in four dimensions thus it cannot give any non-trivial gravitational dynamics in $D\leq 4$, where $D$ is the dimensions of the spacetime. In \cite{Glavan:2019inb}, Glavan and Lin try to extend this theory to four dimensions by rescaling the coupling constant of the Gauss-Bonnet term by a factor of $1/(D-4)$ and taking the limit $D\rightarrow4$. Remarkably, from their point of view, this theory can bypass the Lovelock's theorem and be free from Ostrogradsky instability. Interestingly, its black hole solution was also found in other contexts, such as gravity with conformal anomaly\cite{Cai:2009ua,Cai:2014jea} and quantum corrections\cite{Cognola:2013fva}. This processing method in \cite{Glavan:2019inb} has led to a great deal of discussion, see e.g. \cite{Guo:2020zmf,Fernandes:2020rpa,Anastasiou:2020zwc,Kumar:2020owy,Zhang:2020qam,Arrechea:2020evj,Yang:2020jno,Kumar:2020uyz,Konoplya:2020cbv,Zhang:2020sjh,Mahapatra:2020rds,Gurses:2020ofy,EslamPanah:2020hoj,Arrechea:2020gjw,Gurses:2020rxb,Shu:2020cjw}.

In the present work, we investigate the Hawking evaporation process of the static spherically symmetric neutral AdS black hols in Einstein-Gauss-Bonnet gravity. The thermodynamical properties of these black holes in $ D\geqslant5 $ and $D=4$ are provided in \cite{Cai:2001dz,Fernandes:2020rpa}. Unlike the Schwarzschild-AdS case, which always has a divergent temperature in the late period of the evaporation process\cite{Hawking:1982dh}, the Einstein-Gauss-Bonnet theory reveals richer phenomena by choosing different coupling constant. By testing the Hawking evaporation processes of different Einstein-Gauss-Bonnet AdS black holes, we give some clues to constrain the choice of coupling constant. Especially, the extreme behavior of the cases with negative coupling constant also resembles the Garfinkle-Horowitz-Strominger dilation black hole\cite{Ong:2019glf}.

This paper is organized as follows: In Sec.\ref{ii}, we briefly review the thermodynamics of $ D\geqslant5 $-dimensional neutral Einstein-Gauss-Bonnet AdS black holes. By investigating the Hawking evaporation process of the $D=5$ and $D\geqslant6$ cases, we show that the positive coupling constant may help to solve divergent temperature, while the negative one may make this problem worse. In Sec.\ref{iii}, our investigation is extended to the $D=4$ cases. Summary remarks are given in Sec.\ref{iiii}. We adopt the natural unit system, setting the speed of light in vacuum $c$, the gravitational constant $G_N$, the Planck constant $h$, and the Boltzmann constant $k_B$ equal to one.

\section{Evaporation of Einstein-Gauss-Bonnet AdS Black Holes in $ D\geqslant5 $ }\label{ii}

In this section, first, we will give a brief review of Einstein-Gauss-Bonnet AdS black holes for $D\geqslant5$ dimensions, then we investigate the black hole evaporation process.

\subsection{Black Hole Thermodynamics in $ D\geqslant5 $ Einstein-Gauss-Bonnet AdS spacetimes}

For the $D\geqslant5$ dimensional AdS spacetime, the Einstein-Hilbert action with a Gauss-Bonnet term can be wriiten as
\begin{equation}
	\label{eq1}
	S=\frac{1}{16\pi }\int d^Dx\sqrt{-g}\left(\mathcal{R} +\frac{(D-1)(D-2)}{\ell^2}
	+ \alpha_{\text{GB}}\mathcal{L}_{\text{GB}}\right),
\end{equation}
where  $\alpha_{\text{GB}}$  is the Gauss-Bonnet coupling coefficient with dimension $(length)^2$ 
and the Gauss-Bonnet term $\mathcal{L}_{\text{GB}}=\mathcal{R}_{\mu\nu\gamma\delta}
\mathcal{R}^{\mu\nu\gamma\delta}-4\mathcal{R}_{\mu\nu}\mathcal{R}^{\mu\nu}+\mathcal{R}^{2}$. The Gauss-Bonnet coefficient $\alpha_{\text{GB}}$ is supposed to be a postive value in string theory\cite{Boulware:1985wk}, but for generality we consider both cases of $\alpha_{\text{GB}}>0$ and $\alpha_{\text{GB}}<0$ in the present work. The static spherically symmetric black hole metric in $D$ dimensions can be written in the form of
\begin{equation}
	\label{eq2}
	ds^2 = -f(r)dt^2 +f^{-1}(r)dr^2 +r^2 d\Omega_{D-2}^2,
\end{equation}
with the the metric function\cite{Cai:2001dz}
\begin{equation}
	\label{eq3}
	f(r)=1 +\frac{r^2}{2\alpha}\left ( 1 -
	\sqrt{1+\frac{64 \pi M \alpha }{(D-2)\Omega_{D-2} r^{D-1}}-
		\frac{4\alpha}{\ell^2}} \right), 
\end{equation}
where $\alpha=(D-3)(D-4)\alpha_{\text{GB}}$ and the parameter $M$ is the gravitational mass of the solution. The $\Omega_{D-2}=\frac{(D-1)\pi^{D/2}}{\Gamma(\frac{D+1}{2})}$ is the area of an $(D-1)$-dimensional sphere with unit radius. When $\alpha\rightarrow0$, black hole solution reduces to the Schwarzschild AdS case in \cite{Birmingham:1998nr}. In the limit $M\rightarrow0$, one can obtain the vacuum solution $f(r)=1 +r^2/\ell^2_{eff}$ with the effective AdS radius $\ell_{eff}^2=\frac{\ell^2}{2}\left(1 +\sqrt{1-\frac{4\alpha}{\ell^2}}\right) $, which implies the coupling constant $\alpha$ must obey $\alpha\leqslant \ell^2/4$, beyond that this theory is undefined. The event horizon of the black hole is determined by $f(r_+)=0$, and the mass of black hole $M$ can be obtained as \cite{Cai:2001dz}
\begin{equation}
	\label{eq4}
	M=\frac{(D-2)\Omega_{D-2} r_+^{D-3}}{16\pi }\left (1 +\frac{\alpha }
	{r_+^2} +\frac{r_+^2}{l^2}\right),
\end{equation}
and the Hawking temperature of the black hole can be written as
\begin{equation}
	\label{eq5}
	T =\left. \frac{f'(r)}{4\pi}\right |_{r=r_+} 
	= \frac{(D-1)r_+^4 +(D-3)\ell^2 r_+^2 +(D-5)\alpha \ell^2}
	{4\pi \ell^2 r_+(r_+^2+2\alpha )}.
\end{equation}
In $\alpha>0$, it is worth noting that there is a mass gap $M_c\equiv M(r_+\rightarrow 0)=3\pi\alpha/8$ in $D=5$ that all black holes must satisfy $M\geqslant M_c$, while in $D\geqslant 6$, there exists no mass gap. We can also find the last term in \eqref{eq5} vanishes when $D=5$. Consequently, the temperature $T$ vanishes as $r_+ \rightarrow0$ when $D=5$, while it becomes divergent as $r_+ \rightarrow0$ in higher dimensions. Among various choices for the spacetime dimension $D$, the particular case $D=5$ is qualitatively different from other choices. This is perhaps a consequence of $D=5$ is the lowest dimension in which the Gauss-Bonnet term can affect the local geometry. In $\alpha<0$, there is a extreme horizon radius $r_{min}=\sqrt{2|\alpha|}$ corresponding to a divergent temperature, below that the theory is undefined. The black hole entropy can also be calculated by using the first law of black hole thermodynamics, as well as other thermodynamical quantities. We omit them here.

\subsection{Black Hole Evaporation in $ D\geqslant5 $ Einstein-Gauss-Bonnet AdS spacetimes}

Since we have obtained all the necessary thermodynamical quantities, we are ready to investigate the evaporation of Einstein-Gauss-Bonnet AdS black holes. By imposing an absorbing boundary condition, the black hole mass $M$ should be monotonically-decreasing functions of the time $t$ because of the Hawking radiation. Applying geometric optics approximation, we assume all the emitted massless particles move along null geodesics. If we orient the extra $(D-3)$ angular coordinates in $\mathrm{d}\Omega_{D-2}^{2}$ and normalize the affine parameter $\lambda$, the geodesic equation of the massless particles reads
\begin{align}
	\bigg(\frac{\mathrm{d}r}{\mathrm{d}\lambda}\bigg)^2=E^2-J^2\frac{f(r)}{r^2},
\end{align}
where $E=f(r)\frac{\mathrm{d}t}{\mathrm{d}\lambda}$ is the energy and $J=r^2\frac{\mathrm{d}\theta}{\mathrm{d}\lambda}$ is angular momentum of the emitted particle. Consider an emitted particle coming from just outside the black hole horizon, this particle cannot be detected by the observer on the AdS boundary  if there is a turning point satisfying $\big(\frac{\mathrm{d}r}{\mathrm{d}\lambda}\big)^2=0$. Defining an impact parameter $b\equiv {J}/{E}$, the massless particle can reach infinity only if 
\begin{align}
	\frac{1}{b^2}> \frac{f(r)}{r^2},
\end{align}
for all $r> r_+$. The impact factor $b_c$ is defined by the maximal value of ${f(r)}/{r^2}$, which depends on the exact form of metric function ${f(r)}$.

If we obtain the impact factor $b_c$, we can get the Hawking emission rate in $D$-dimensional spacetime according to Stefan-Boltzmann law, yields\cite{Vos1989,Cardoso:2005cd,Xu:2019wak,Xu:2019krv,Xu:2020xsl}
\begin{align}
\frac{\mathrm{d} M}{\mathrm{d}t}=- C b_c^{D-2} T^D,
\label{law}
\end{align}
with the constant $C=(D-2)\pi^{\frac{D}{2}-1}\Omega_{D-2}\frac{k^D}{h^{D-1}c^{D-1}}\frac{\Gamma(D)}{\Gamma(\frac{D}{2})}\zeta(D)$. Without loss of generality, we set the constant $C=1$. The Stefan-Boltzmann law implies that in $D$-dimensional spacetime the emission power is proportional to the $(D-2)$-dimensional cross section $b_c^{D-2}$ and the $(D-1)$-dimensional space (spatial dimension only) photon energy $T^D$. Notice that the photon energy term $T^D$ has the higher order, so the behavior of temperature $T$, especially the asymptotical behavior, will play a leading role in black hole evaporation process. Next, we are ready to study the black hole evaporation for various features of $T$ and $b_c$.

Using scaling analysis we know $M\sim l^{D-3}$, $T\sim l^{-1}$, $b_c\sim l$, $\ell\sim l$ and $\alpha\sim l^2$, where $l$ has the dimension $(length)$. In order to get dimensionless parts of $M$, $T$ and $b_c$, by defining dimensionless variables $x\equiv {r_+}/{\ell}$ and $y\equiv {\alpha}/{\ell^2}$, we can express $M$ and $T$ as
\begin{equation}
	\label{eq7}
	M=\frac{(D-2)\Omega_{D-2} }{16\pi }\left (1 +\frac{y}{x^2} +x^2 \right) x^{D-3}\ell^{D-3},
\end{equation}
\begin{equation}
	\label{eq8}
	T = \frac{(D-1)x^4 +(D-3)x^2 +(D-5)y}
	{4\pi  x(x^2+2y )}\ell^{-1}.
\end{equation}
The $b_c$ depends on the exact form of metric function $f(r)$, so we cannot get the universal solution for various dimensions. 

\begin{figure}[h!]
	\begin{center}
		\includegraphics[width=0.40\textwidth]{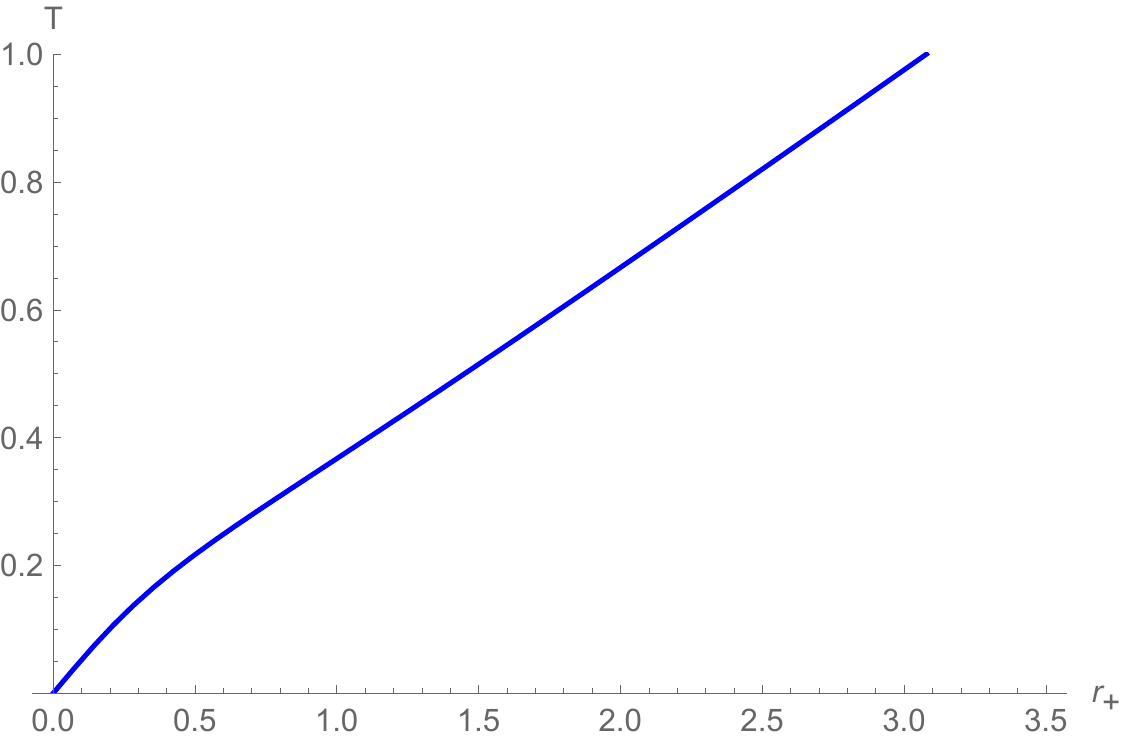}
		\includegraphics[width=0.40\textwidth]{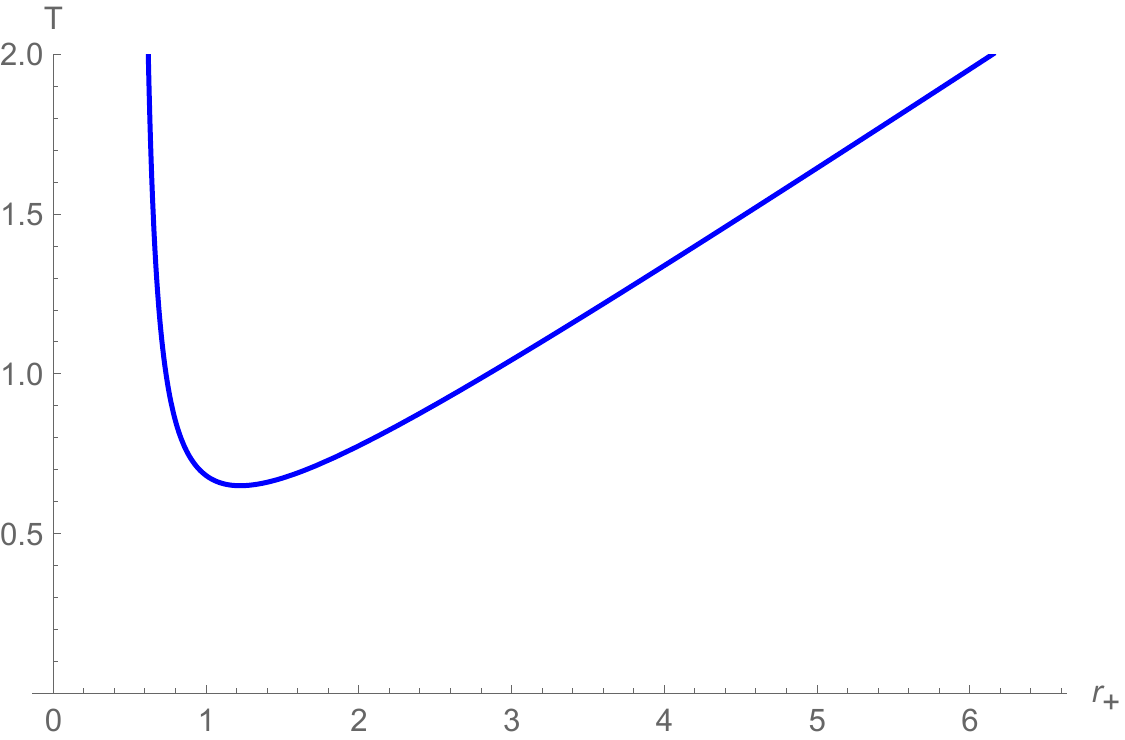}
		\vspace{-1mm}
		\caption{The temperature of black holes in $5$ dimensions with $\alpha>0$ and $\alpha<0$ respectively. The left figure we set $\alpha=0.15$, $\ell=1$, while in the right figure $\alpha=-0.15$, $\ell=1$. }
		\label{fig1}
	\end{center}
\end{figure}

Firstly we start from the case of $D=5$, which is qualitatively different from the cases of  $D\geqslant6$. When $D=5$, the temperature \eqref{eq5} can be written as
\begin{equation}
	\label{eq9}
	T = \frac{2 r_+^3+\ell^2 r_+}{2\pi \ell^2(r_+^2+2\alpha) }.
\end{equation}
In FIG.\ref{fig1} we present the temperature of black holes in $5$ dimensions with $\alpha>0$ and $\alpha<0$ as the function of $r_+$ respectively. For $\alpha>0$, the temperature starts from zero at $ r_+=0$ and goes to infinity when $r_+\rightarrow\infty$. The behavior of temperature in $\alpha<0$ is quite different from the case of $\alpha>0$. There is an extreme horizon radius $r_{min}=\sqrt{2|\alpha|}$, which coincides with an additional singularity. As shown in the figure, the temperature is divergent at $r_{min}$ when $\alpha<0$, and as a result, the thermodynamical properties of case $\alpha<0$ resembles the case of $\alpha\rightarrow0$ that without Gauss-Bonnet term.

In order to get the impact factor $b_c$, solving the equation $\frac{\partial}{\partial r}\frac{f(r)}{r^2}=0$, we can find the root
\begin{equation}
	\label{eq10}
	r_p=\frac{2}{\sqrt{3\pi}}\left(\frac{6\pi \ell^2 M \alpha-16 \ell^2 M^2}{4\alpha-\ell^2}\right)^{1/4}.
\end{equation}
This radius corresponds to the unstable photon orbit, and the impact factor is given by $b_c= {r_p}/{\sqrt{f(r_p)}}$. Similarly, taking $x\equiv {r_+}/{\ell}$ and $y\equiv {\alpha}/{\ell^2}$, one can obtain the $b_c$ as
\begin{equation}
	\label{eq11}
	b_c=\left( \frac{\sqrt{4y-1}}{2\sqrt{-(x^2+x^4)(x^2+x^4+y)}}-\frac{-1+\sqrt{-\frac{(x^2+x^4+y)(4y-1)}{x^2+x^4}}}{2y} \right) ^{-\frac{1}{2}}\ell
\end{equation}
For $D=5$, inserting the black hole mass $M$ \eqref{eq7}, temperature $T$ \eqref{eq8} and impact factor $b_c$ \eqref{eq11} into Stefan-Boltzmann law, we can obtain
\begin{align}
	\mathrm{d}t=\ell^{4}F_1(x,y)\mathrm{d}x,
\end{align}
where $F_1(x,y)$ is a complicated function which is not worth explicitly written. Setting $y$ to be a constant and integrating the equation from $\infty$ to $x_{min}$, where $x_{min}=0$ for $\alpha>0$ and $x_{min}=r_{min}/\ell=\sqrt{2|y|}$ for $\alpha<0$, one can obtain the black hole lifetime is divergent in $\alpha>0$ and of the order $\ell^{4}$ in $\alpha<0$. 

\begin{figure}[h!]
	\begin{center}
		\includegraphics[width=0.40\textwidth]{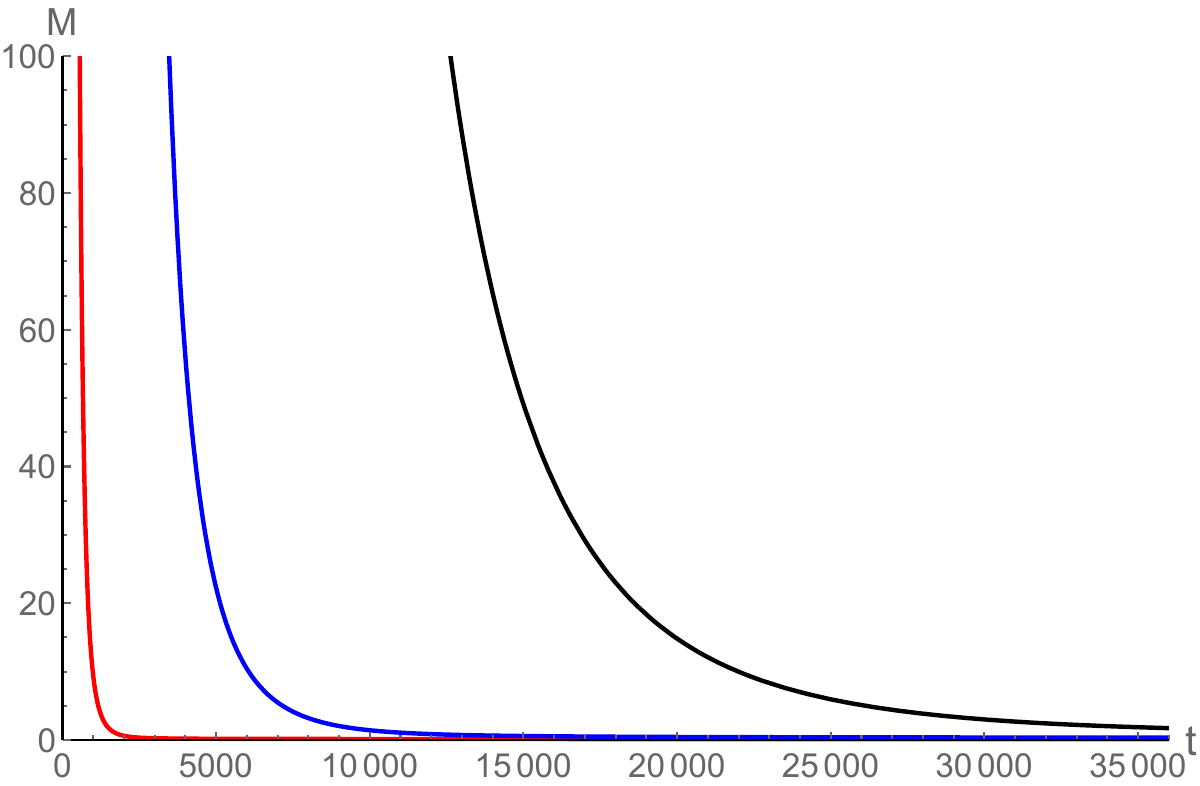}
		\includegraphics[width=0.40\textwidth]{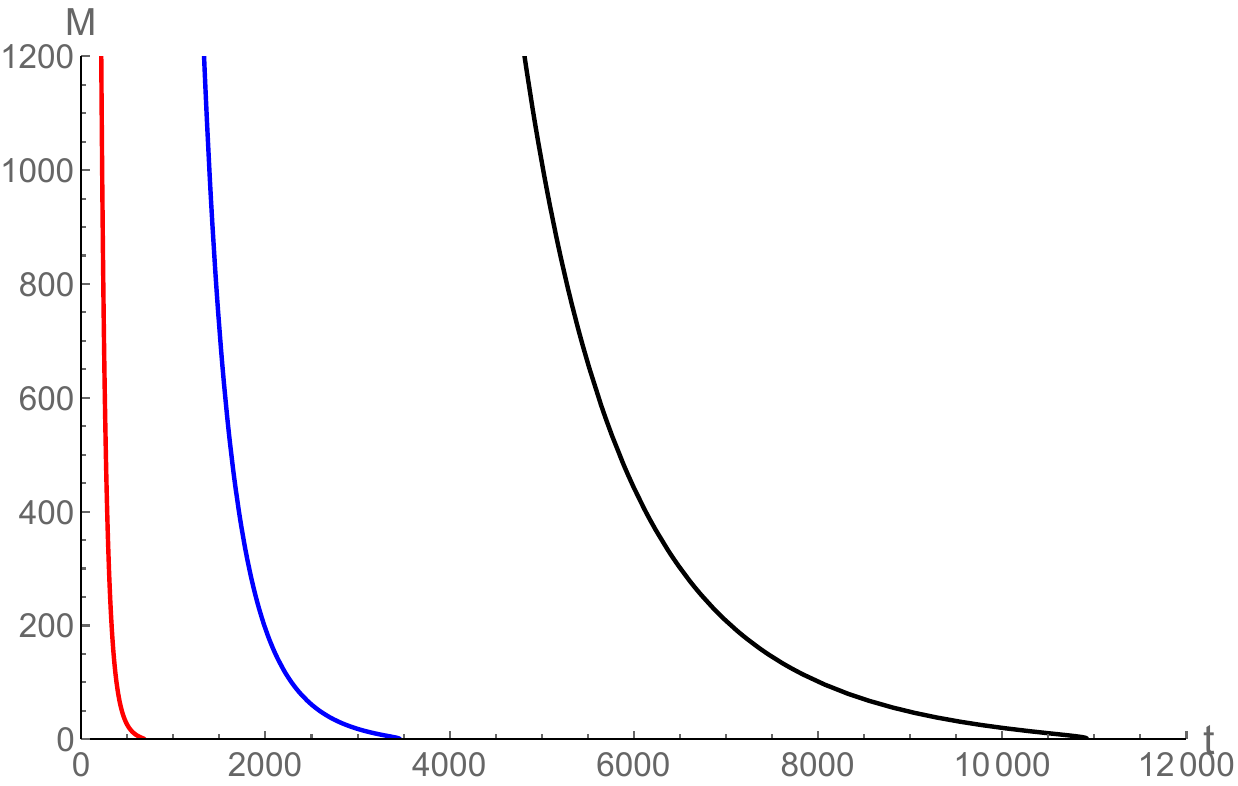}
		\vspace{-1mm}
		\caption{The numerical results of the black hole mass $M$ as function of the lifetime $t$ with setting $y>0$ and $y<0$ (corresponding to $\alpha>0$ and $\alpha<0$, respectively). In the left figure, we set $y=0.1$ with $\ell=1$, $\ell=1.5$ and $\ell=2$ from left to right respectively. Note that there is a mass gap $M_c=\frac{3\pi \alpha}{8}$ for $y>0$ ($\alpha>0$). In the right figure, we set $y=-0.1$ with $\ell=1$, $\ell=1.5$ and $\ell=2$ from left to right, corresponding to minimum mass $M_{min1}=0.164934$, $M_{min2}=0.371101$ and $M_{min3}=0.659734$, respectively. }
		\label{fig2}
	\end{center}
\end{figure}

In FIG.\ref{fig2}, we plot some numerical results of the black hole mass $M$ as function of the lifetime $t$ by setting $y>0$ and $y<0$ (corresponding to $\alpha>0$ and $\alpha<0$ respectively). In the case of $\alpha>0$, we set $y=0.1$, and the initial mass is taken to infinity. The black holes lose their mass in a ``short'' time, but the evaporation gets harder when the black hole mass and temperature become smaller near the $T=0$ state, thus the black hole will have an infinite lifetime, satisfying the third law of black hole thermodynamics. This means that the black hole becomes a remnant, which may help us to ameliorate the information paradox \cite{Chen:2014jwq}. In the case $\alpha<0$, there is an extreme horizon radius at $r_{min}=\sqrt{2|\alpha|}$, where the temperature of black hole is divergent coinciding with an additional singularity \cite{Cai:2001dz}. This $r_{min}$ corresponds to a mass $M_{min}=M(r_{min})$, and the black hole will take a finite time to decrease the mass to  $M_{min}$. Interestingly, this phenomenon resembles the extreme case of Garfinkle-Horowitz-Strominger dilation black hole\cite{Ong:2019glf}.

Now we extend our study to $D\geqslant6$. The temperature behavior in $D\geqslant6$ is quite different from the case $D=5$ when $\alpha>0$. For $D\geqslant6$, the temperature is divergent at $r_+=0$ and goes to infinity again when $r_+\rightarrow\infty$\cite{Cai:2001dz}, which resembles the case of Schwarzschild AdS. On the other hand, the temperature behavior of case $\alpha<0$ in $D\geqslant6$ spacetimes is similarly with the case $\alpha<0$ in 5 dimensions, where there is also the smallest horizon radius at $r_{min}=\sqrt{2|\alpha|}$. Although $\frac{\partial}{\partial r}\frac{f(r)}{r^2}=0$ is a equation of higher order in $D\geqslant6$ dimensions and we cannot get the explicit form of $b_c$, we can still conclude $b_c \sim \ell$ by scaling analysis. As we have emphasized earlier in the present work, the photon energy term $T^D$  plays the leading role in the black hole evaporation process. Similarly, inserting the black hole mass $M\sim \ell^{D-3}$, temperature $T\sim\ell^{-1}$ and impact factor $b_c\sim \ell$ into Stefan-Boltzmann law, we can also obtain the $D$-dimensional Stefan-Boltzmann law as
\begin{align}
	\mathrm{d}t=\ell^{D-1}F_2(x,y)\mathrm{d}x,
\end{align}
where $F_2(x,y)$ is still a complicated function. We can conclude this evaporation process qualitatively by analyzing the temperature asymptotic behavior of the black hole. When $\alpha>0$, the temperature $T$ is divergent in both $r_+\rightarrow \infty$ and $r_+\rightarrow 0$, which means that $\int^{0}_{\infty} F_2(x,y)\mathrm{d}x$ is convergent and just depends on the value of $y$. Hence, the lifetime of the black hole will be a finite value of the order $\ell^{D-1}$ with any arbitrarily large initial mass, which resembles the Schwarzschild AdS black hole in Einstein gravity \cite{Page:2015rxa}. For $\alpha<0$, the temperature of black hole is still divergent at $r_{min}=\sqrt{2|\alpha|}$, thus the qualitative features of the evaporation process in $D\geqslant6$ dimensions are consistent with the case of $\alpha<0$ in $5$ dimensions.

\section{Evaporation of novel $ 4D $ Einstein-Gauss-Bonnet AdS Black Holes}\label{iii}

Since Gauss-Bonnet term is a topological invariant in four dimensions, previously the investigation of Gauss-Bonnet gravity is limited to $D\geq 5$ dimensions. This situation is changed when Glavan \& Lin declare that rescaling the coupling coefficient $\alpha_{\text{GB}}$ of the Gauss-Bonnet term by a factor of $1/(D-4)$ can lead to non-trivial dynamics when taking the limit $D\rightarrow4$, thus extend the classic Einstein-Gauss-Bonnet theory to $4$ dimensions \cite{Glavan:2019inb}. Remarkably, in their point of view, this theory can bypass Lovelock’s theorem and be free from Ostrogradsky instability. The black hole solution in AdS space has been provided in \cite{Fernandes:2020rpa}. It is worth noting that this black hole solution has been found in other backgrounds, such as gravity with conformal anomaly \cite{Cai:2009ua,Cai:2014jea} and with quantum corrections\cite{Cognola:2013fva}. In this section, we will take the metric in \cite{Fernandes:2020rpa} and investigate the qualitative evaporation process. We hope our result may help inspiring the researches on $4D$ Einstein-Gauss-Bonnet gravity. 

\subsection{Black Hole Thermodynamics in $ 4D $ Einstein-Gauss-Bonnet AdS spacetimes}
In this section we give a brief review on thermodynamics of $4D$  Einstein-Gauss-Bonnet AdS black holes. The static spherically symmetric balck hole solution takes the form
\begin{equation}
	ds^2 = -f(r)dt^2 +f^{-1}(r)dr^2 +r^2 d\Omega_2^2
	\label{eq12}
\end{equation}
with the the metric function\cite{Fernandes:2020rpa}
\begin{equation}
		f(r)=1+\frac{r^2}{2\alpha}\left(1- \sqrt{1+4\alpha\left(\frac{2M}{r^3}-\frac{1}{\ell^2}\right)}\right),
	\label{eq13}
\end{equation}
where $\alpha$ is coupling constant and the parameter $M$ is the mass of black hole. Similarly, in the vanishing limit of mass $M\rightarrow0$, one can obtain the vacuum solution $f(r)=1 +r^2/\ell^2_{eff}$ with the effective AdS radius $\ell_{eff}^2=\frac{\ell^2}{2}\left(1 +\sqrt{1-\frac{4\alpha}{\ell^2}}\right) $, thus the couping constant $\alpha$ must obey $\alpha\leqslant \ell^2/4$. The solution still keeps asymptotically AdS when $\alpha<0$. Notably, the black hole solution will recover the solution in Einstein gravity when $\alpha\rightarrow0$. Solving $f(r_+)=0$ at the event horizon $r_+$, we can obtain the mass of black hole
\begin{equation}
	M=\frac{r_+}{2}\left(1+\frac{r_+^2}{\ell^2} + \frac{\alpha}{r_+^2} \right).
	\label{eq14}
\end{equation}
The Hawking temperature of the black hole can be obtain as
\begin{equation}
	T=\left. \frac{f'(r)}{4\pi}\right |_{r=r_+} =\frac{3r_+^4 + \ell^2(r_+^2-\alpha)}{4\pi \ell^2 r_+ (r_+^2 + 2\alpha)},
	\label{eq15}
\end{equation}
For $\alpha>0$, solving $T=0$, there is a critical horizon radius $r_c=\sqrt{\left(-\ell^2+ \sqrt{\ell^2(\ell^2+12\alpha)}\right)/6}$ corresponding to a vanishing temperature. Similarly there will also be a critical mass 
\begin{equation}
	M_c=\frac{-\ell^2+12\alpha+\sqrt{\ell^4+12\ell^2\alpha}}{3\sqrt{6}\sqrt{-\ell^2+\sqrt{\ell^4+12\ell^2\alpha}}},
\end{equation}
which depends on the AdS radius $\ell$ and the coupling constant $\alpha$. For $M<M_c$, there is no black hole. 
The black hole entropy can also be calculated by the first law of black hole thermodynamics, see \cite{Fernandes:2020rpa}.

\subsection{Black Hole Evaporation in $4D$ Einstein-Gauss-Bonnet AdS spacetimes }

\begin{figure}[h!]
	\begin{center}
		\includegraphics[width=0.40\textwidth]{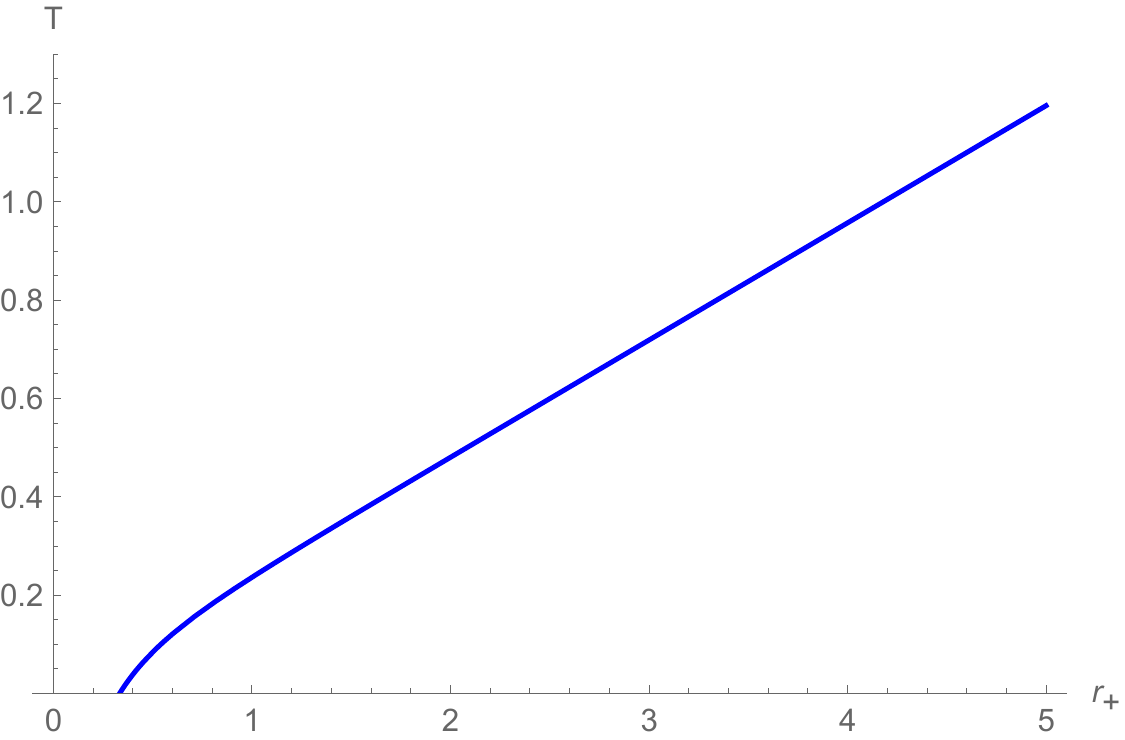}
		\includegraphics[width=0.40\textwidth]{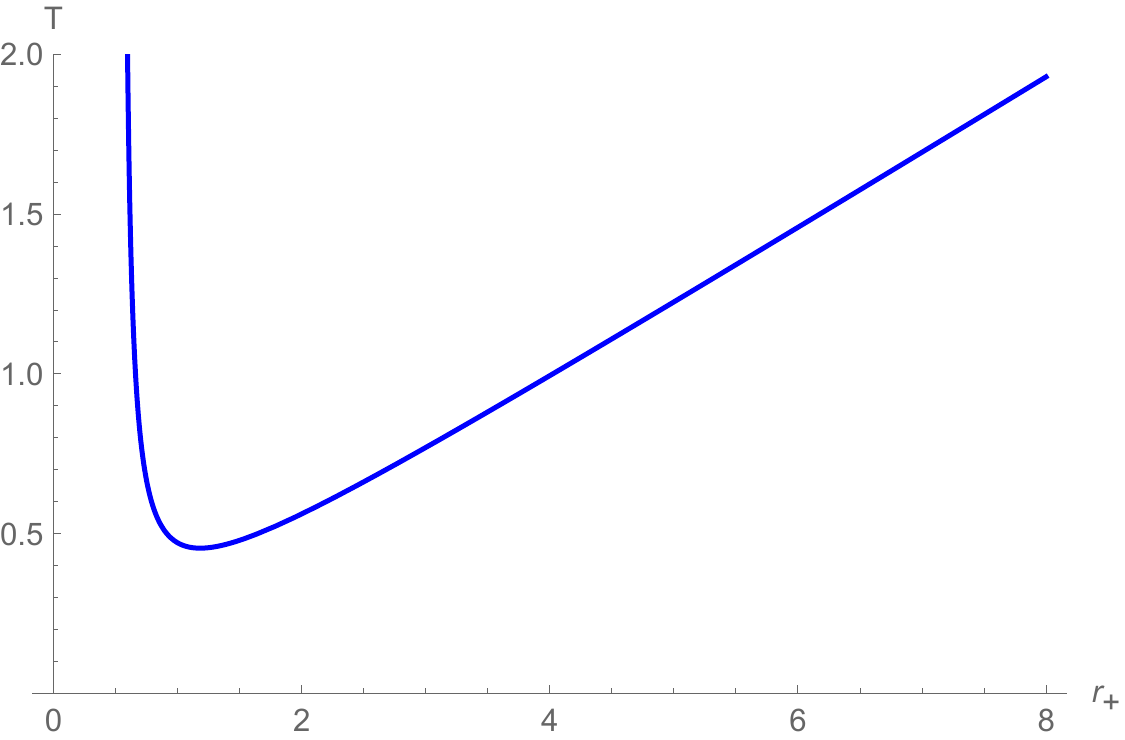}
		\vspace{-1mm}
		\caption{The temperature of black holes in $4$-dimensions with $\alpha>0$ and $\alpha<0$ respectively. The left figure we set $\alpha=0.15$, $\ell=1$, while in the right $\alpha=-0.15$, $\ell=1$. }
		\label{fig3}
	\end{center}
\end{figure}

After the review of the thermodynamics of $ 4D $ Einstein-Gauss-Bonnet AdS black holes, we are ready to explore the black holes evaporation. Imposing the same assumptions that we take in the case $D \geqslant5$, we can obatin the Stefan-Boltzmann law in $4D$, reads
\begin{align}
	\frac{\mathrm{d} M}{\mathrm{d}t}=-  b_c^{2} T^4.
\end{align}
Note that the behavior of temperature $T$ still plays a leading role in black hole evaporation process since it occupies the highest order. In order to analyze the qualitative behavior of the black hole temperature $T$, we provide the examples with $\alpha>0$ and $\alpha<0$ in FIG.\ref{fig3}. It is clear that $T=0$ at a critical horizon radius $r_c=\sqrt{(-\ell^2+\sqrt{\ell^2(\ell^2+12\alpha)})/6}$, which means that exists a critical mass $M_c$. When $\alpha<0$, this situation resembles the case $\alpha<0$ in $ D\geqslant5 $ which also have divergent temperature at $r_{min}=\sqrt{2|\alpha|}$.

Solving the equation $\frac{\partial}{\partial r}\frac{f(r)}{r^2}=0$, we can obtain the root corresponding to the photon orbit as
\begin{align}
	r_p=\left( -\frac{4M\alpha}{1-\frac{4\alpha}{\ell^2}}-\sqrt{-\frac{27M^6}{(1-\frac{4\alpha}{\ell^2})^3}+\frac{16M^2\alpha^2}{(1-\frac{4\alpha}{\ell^2})^2}}\right)^{\frac{1}{3}}+\left( -\frac{4M\alpha}{1-\frac{4\alpha}{\ell^2}}+\sqrt{-\frac{27M^6}{(1-\frac{4\alpha}{\ell^2})^3}+\frac{16M^2\alpha^2}{(1-\frac{4\alpha}{\ell^2})^2}}\right) ^{\frac{1}{3}}.
\end{align}
Similarly, using the dimensionless variables $x\equiv {r_+}/{\ell}$ and $y\equiv {\alpha}/{\ell^2}$, we can re-express the black hole mass \eqref{eq14} and the temperature \eqref{eq15} as
\begin{align}
	M &=\frac{x^2+x^4+y}{2x}\ell,\\
	T &= \frac{x^2+3x^4-y}{4\pi x^3+8 \pi xy}\ell^{-1}.
\end{align}
The impact factor $b_c= {r_p}/{\sqrt{f(r_p)}}$ can also be expressed in the form of 
\begin{align}
	b_c&=\left( \frac{4}{\left(2(A-B)^{1/3}+2(A+B)^{1/3} \right) ^2}-\frac{-1+\sqrt{1-4y+\frac{32y(x^2+x^4+y)}{x\left(2(A-B)^{1/3}+2(A+B)^{1/3} \right) ^3}}}{2y}\right) ^{-\frac{1}{2}}\ell\nonumber \\
	A&=\frac{2  y (x^2 + x^4 + y)}{x (-1 + 4 y)},\quad B=\frac{1}{8} \sqrt{\frac{ (x^2 + x^4 + y)^2 (256 x^4 (1 - 4 y) y^2 - 
	27 (x^2 + x^4 + y)^4)}{x^6 (1 - 4 y)^3}}.
\end{align}

\begin{figure}[h!]
	\begin{center}
		\includegraphics[width=0.40\textwidth]{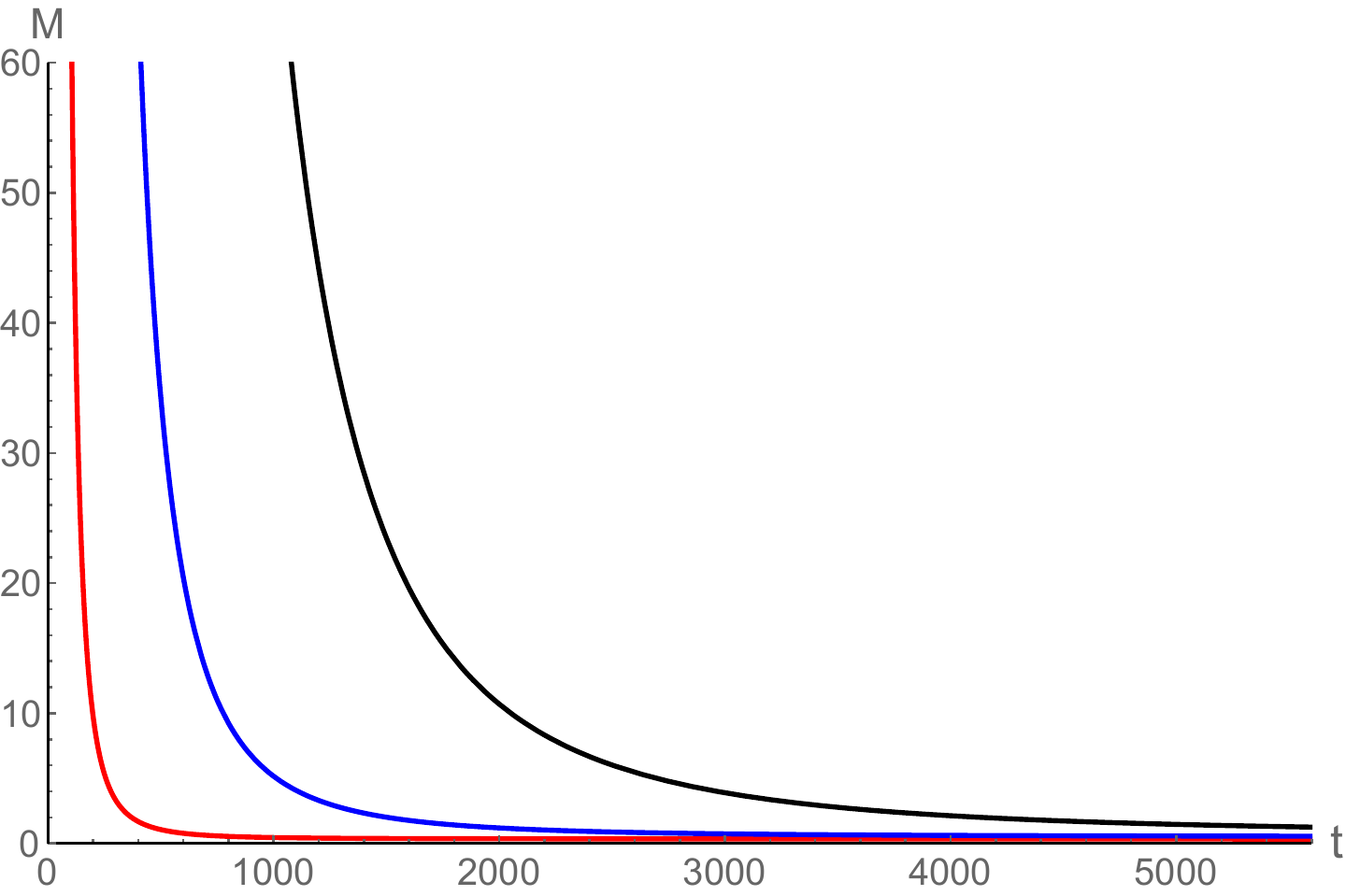}
		\includegraphics[width=0.40\textwidth]{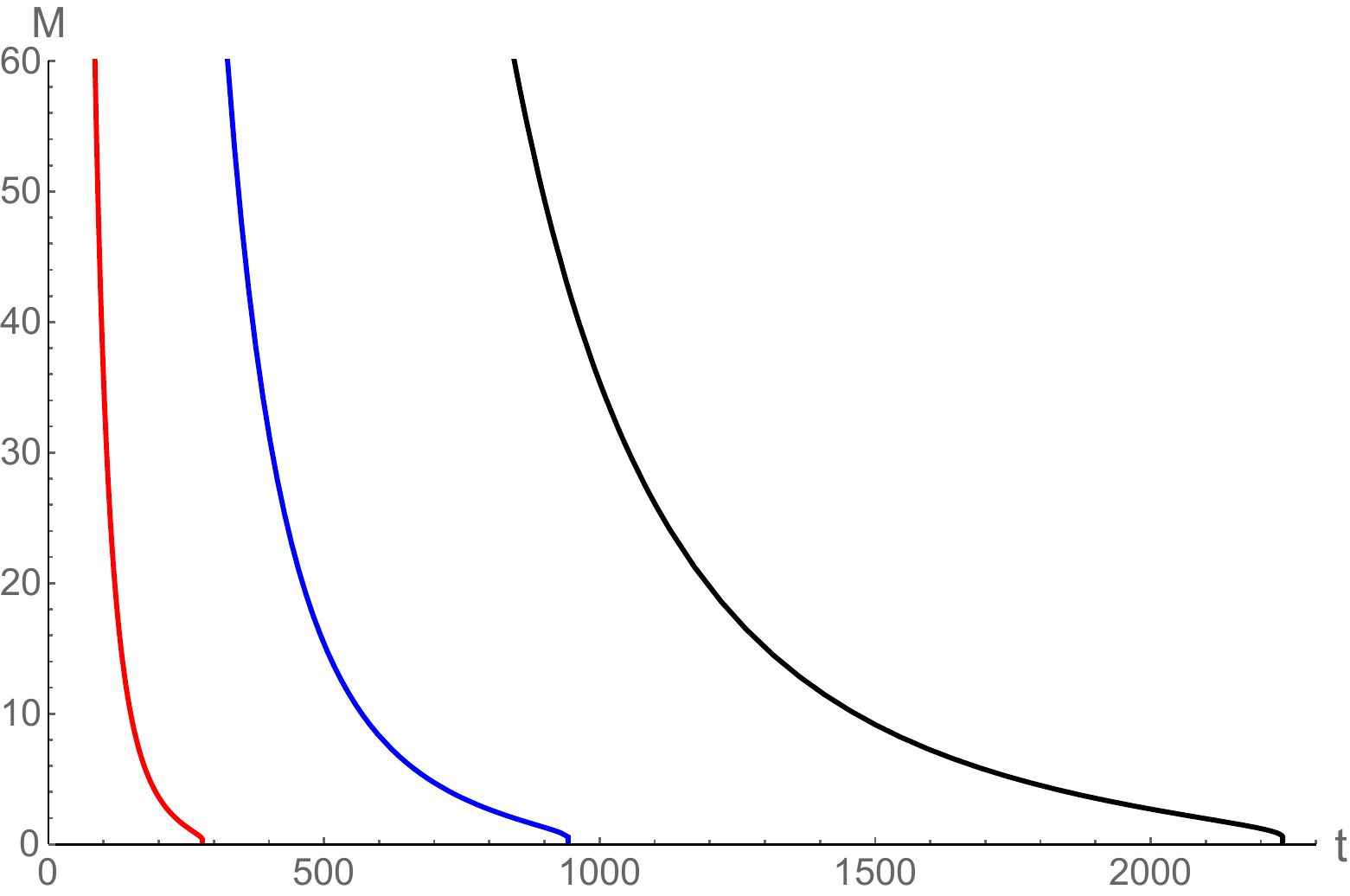}
		\vspace{-1mm}
		\caption{The numerical results of the $ 4D $ Einstein-Gauss-Bonnet AdS black hole mass $M$ as function of the lifetime $t$ with setting $y>0$ and $y<0$ (corresponding to $\alpha>0$ and $\alpha<0$, respectively). In the left figure, we set $y=0.1$ with $\ell=1$, $\ell=1.5$ and $\ell=2$ from left to right, corresponding to critical mass $M_{c1}=0.329509$, $M_{c2}=0.494264$ and $M_{c3}=0.659019$, respectively. In the right figure, we set $y=-0.1$ with $\ell=1$, $\ell=1.5$ and $\ell=2$ from left to right, corresponding to minimum mass $M_{min1}=0.156525$, $M_{min2}=0.234787$ and $M_{min3}=0.313050$, respectively. }
		\label{fig4}
	\end{center}
\end{figure}

Inserting the above $M$, $T$ and $b_c$ into the Stefan-Boltzmann law in $4D$ spacetimes, we can obtain the lifetime relationship of black hole evaporation, reads
\begin{align}
	\mathrm{d}t=\ell^{3}F_3(x,y)\mathrm{d}x,
\end{align}
where $F_3(x,y)$ is a complex function related to the $M$, $T$ and $b_c$. Similarly, integrating this equation from $\infty$ to $x_{min}$ (where the temperature $T$ has divergence or $T=0$), one can find that the lifetime is of the order $\sim \ell^3$ with $\alpha<0$ or divergent with $\alpha>0$, which resembles the case in $D\geqslant5$ dimensions. We plot some numerical results of the $ 4D $ Einstein-Gauss-Bonnet AdS black hole in FIG.\ref{fig4}. As shown in the figure, the case of $\alpha>0$ is very similar to the case of $5D$ with $\alpha>0$, just has the difference in the value of critical mass which in case $5D$ is $3\pi\alpha/8$, while in case $4D$ equals to $M_c$. While in $\alpha<0$, the qualitative features of the evaporation process are parallel to the case $\alpha<0$ in $5D$, which also possess a minimal horizon radius at $r_{min}=\sqrt{2|\alpha|}$. The qualitative features of the evaporation process identify what implied from the asymptotical behavior of the black hole temperature $T$.

\section{Summary}\label{iiii}

By introducing the higher order curvature term, Einstein-Gauss-Bonnet theory is supposed to solve a class of singularity problem of black holes. Because of the modification, the thermodynamic properties of Einstein-Gauss-Bonnet AdS black holes are quite different from the Schwarzschild-AdS case. In the present work, we investigate the  evaporation process of Einstein-Gauss-Bonnet AdS black holes in $D$-dimensional cases with both positive and negative coupling constant. Firstly, for the case of $D=5$ and $\alpha>0$, the temperature $T$ is not divergent at a small distance and admits the state of $T=0$ at $r_+=0$, which is different from the Schwarzschild-AdS case. Applying the geometric optics approximation and absorbing AdS boundary condition, we investigate the evaporation of black holes by using the $D$-dimensional Stefan-Boltzmann law. The black holes will lose an infinite amount of mass in a ``short'' time, then the evaporation will become slower and slower near the $T=0$ state, which means that the black holes with $\alpha>0$ have a divergent lifetime in $5$-dimensional spacetime. This phenomenon also obeys the third law of black hole thermodynamics. When $\alpha<0$, the temperature is divergent at $r_{min}=\sqrt{2|\alpha|}$, and the black holes will always spend finite time $t\sim \ell^4$ to reach the extreme mass $M(r_{min})$. For the cases of $D\geqslant6$, although it is difficult to get the exact form of impact factor $b_c$, we can still dope out the evaporation process by analyzing the asymptotic behavior of the temperature. Since its temperature is always positive and divergent in $r_+\rightarrow0$ and $r_+\rightarrow \infty$, the black hole lifetime is always of the order $t\sim \ell^{D-1}$, which resembles the Schwarzschild-AdS case. Applying the ``novel'' metric, we extend our investigation to $4D$. Just with a difference in the value of critical mass between case $4D$ and case $5D$ when $\alpha>0$, the evaporation behaviors of both $\alpha>0$ and $\alpha<0$ are parallel to the ones in $5D$.

In the present work, we consider both $\alpha>0$ and $\alpha<0$ in different dimensions and find that some choices of $\alpha$ are not applicable. The case of $\alpha>0$ may help us solving the problem of the divergent temperature in the terminal of the evaporation process of the Schwarzschild-AdS case, while the $\alpha<0$ could make it worse. Also, in the cases $D=4,5$ there always have remnants, while in the cases $D\geqslant6$ that all the black holes will evaporate in a finite time. This phenomenon is quite similar to the Ho\v{r}ava-Lifshitz case\cite{Xu:2020xsl}, which also has the dimensional dependent evaporation behavior.

Finally, we suggest some topics that can be further investigated. In consideration of the lifetime of the black holes in Einstein-Gauss-Bonnet and Ho\v{r}ava-Lifshitz gravity are dimensional dependent, there may be some relationship between the dimensions $D$ and asymptotic behavior of evaporation, finding this relationship may become an interesting work\cite{Emparan:2020inr,Holdt-Sorensen:2019tne}. Besides, this work can be extended to charged black holes in Einstein-Gauss-Bonnet gravity \cite{Torii:2005nh,Fernandes:2020rpa}. By considering the Schwinger effect and the Stefan-Boltzmann law, we could calculate the charged particles emission and the mass loss\cite{Xu:2019wak}. How to describe the evolution of charged black holes in Einstein-Gauss-Bonnet gravity will be challenging and meaningful work. Furthermore, modification to the AdS/CFT correspondence can be considered from the perspective of Einstein-Gauss-Bonnet AdS black hole evaporation. In the present work, our derivation relies on the geometric optics approximation and absorbing boundary condition. Nowadays there have been many papers on evaporating black hole in AdS in connection with the formation of islands \cite{Penington:2019npb,Almheiri:2019psy,Almheiri:2020cfm}. It would be interesting to study the connection among absorbing boundary conditions, black hole evaporation in AdS spacetime, and the islands. We hope to consider these questions in our future work.

\section*{Acknowledgements}

We would like to thank Yen Chin Ong, Rui-Hong Yue and De-Cheng Zou for useful discussion. This work is supported by National Natural Science Foundation of China (NSFC) under grant No. 11575083, 11565017, and the Natural Science
Foundation of the Jiangsu Higher Education Institutions of China (Grant No. 20KJD140001).


\providecommand{\href}[2]{#2}\begingroup
\footnotesize\itemsep=0pt
\providecommand{\eprint}[2][]{\href{http://arxiv.org/abs/#2}{arXiv:#2}}

\end{document}